\documentclass[12pt,epsf]{article}
\usepackage{amsmath}
\usepackage{amsthm, amssymb}
\usepackage{graphicx}
\usepackage{setspace}
\usepackage{subfigure}
\usepackage{cite}

\theoremstyle{remark}

\newcommand{\be}{\begin{equation}}
\newcommand{\ee}{\end{equation}}
\newcommand{\bea}{\begin{eqnarray}}
\newcommand{\eea}{\end{eqnarray}}
\newcommand{\bear}{\begin{eqnarray}}
\newcommand{\eear}{\end{eqnarray}}
\newcommand{\beas}{\begin{eqnarray*}}

\newcommand{\eeas}{\end{eqnarray*}}
\newcommand{\ba}{\begin{array}}
\newcommand{\ea}{\end{array}}

\newcommand{\ra}

\newcommand{\pd}[2][1]{\ifnum#1=1 \frac{\partial}{\partial {#2}} \else
  \frac{\partial^#1}{\partial {#2}^{#1}}\fi}
\newcommand{\dpd}[2][1]{\ifnum#1=1 \dfrac{\partial}{\partial {#2}} \else
  \frac{\partial^#1}{\partial {#2}^{#1}}\fi}
\newcommand{\td}[2][1]{\ifnum#1=1 \frac{d}{d{#2}} \else
  \frac{d^#1}{d{#2}^{#1}}\fi}

\newcommand{\nbox}{{\,\lower0.9pt\vbox{\hrule \hbox{\vrule height 0.2 cm \hskip 0.19 cm \vrule height 0.2 cm}\hrule}\,}}

\def\href#1#2{#2}

\usepackage{color}

\usepackage{xcolor}
\usepackage[citebordercolor=green, linkbordercolor={ 0 0 1}, linktocpage=true]{hyperref}

\newcommand\blfootnote[1]{%
  \begingroup
  \renewcommand\thefootnote{}\footnote{#1}%
  \addtocounter{footnote}{-1}%
  \endgroup
}

\begin{document}
\begin{titlepage}
\begin{NoHyper}
\hfill
\vbox{
    \halign{#\hfil         \cr
           } 
      }  
\vspace*{20mm}
\begin{center}
{\Large \bf No Simple Dual to the Causal Holographic Information?}

\vspace*{15mm}
\vspace*{1mm}
Netta Engelhardt$^{a}$ and Aron C. Wall$^{b}$
\vspace*{1cm}
\blfootnote{nengelhardt@princeton.edu, aroncwall@gmail.com}

{$^{a}$ Department of Physics, Princeton University\\
Princeton, NJ 08544 USA\\
\vspace{0.25cm}
$^{b}$ Institute for Advanced Study,\\
Einstein Drive, Princeton NJ 08540 USA
}

\vspace*{1cm}
\end{center}

\begin{abstract}
In AdS/CFT, the fine grained entropy of a boundary region is dual to the area of an extremal surface $X$ in the bulk.  It has been proposed that the area of a certain `causal surface' $C$---i.e. the `causal holographic information' (CHI)---corresponds to some coarse-grained entropy in the boundary theory.  We construct two kinds of counterexamples that rule out various possible duals, using (1)~vacuum rigidity and (2)~thermal quenches.  This includes the `one-point entropy' proposed by Kelly and Wall, and a large class of related procedures.  Also, any coarse-graining that fixes the geometry of the bulk `causal wedge' bounded by $C$, fails to reproduce CHI.  This is in sharp contrast to the holographic entanglement entropy, where the area of the extremal surface $X$ measures the same information that is found in the `entanglement wedge' bounded by $X$.
\end{abstract}

\end{NoHyper}

\end{titlepage}
\tableofcontents
\vskip 1cm
\begin{spacing}{1.2}

\section{Introduction}
Since the Anti-de Sitter (AdS)/Conformal Field Theory (CFT) correspondence was first proposed~\cite{Mal97, Wit98a, GubKle98}, understanding the dictionary between bulk and boundary has been a primary goal in the field.
We now understand that the density matrix of a boundary subregion ${\cal R}$  is dual to the entanglement wedge~\cite{CzeKar12,Wal12, HeaHub14, JafLew15, DonHar16}, a bulk volume bounded by the the HRT surface~\cite{HubRan07} --- the minimal area extremal surface homologous to ${\cal R}$.

\begin{figure}[t]
\centering
\includegraphics[width=5cm]{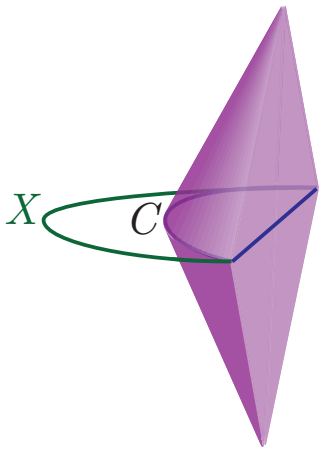}
\caption{The causal wedge, given by the region in past and future causal contact with a boundary subregion. The causal surface $C$ is the intersection of the past and future horizons. The HRT surface X lies outside of the causal wedge and in a spacelike direction~\cite{Wal12}.}
\label{fig:wedges}
\end{figure}

A natural geometric object whose boundary interpretation remains elusive is the causal wedge $C_{W}[{\cal R}]$ of a boundary subregion ${\cal R}$, defined in~\cite{BouLei12, HubRan12} as the bulk region that can both send and receive signals from the boundary domain of dependence $D[{\cal R}]$ of ${\cal R}$. See Fig.~\ref{fig:wedges}.

Various prescriptions have been proposed for reconstructing the bulk metric and other bulk fields within the causal wedge~\cite{HamKab06, Kab11, Hee12, KabLif12, KabLif13b, BouFre12, Mor14, EngHor16a, EngHor16c} from the dual field theory. There has been a particular emphasis on developing an understanding of the causal surface: the outer rim of the causal wedge, i.e. the intersection of the past and future horizons of $D[{\cal R}]$ (the surface labeled $C$ in Fig.~\ref{fig:wedges}). Hawking's area increase theorem~\cite{Haw71} implies that the area of the causal surface is the smallest cross-section of the boundary of the causal wedge.

It has been argued that the area of the causal surface, termed the causal holographic information (CHI), has an information-theoretic boundary dual~\cite{HubRan12}. Since CHI of a region ${\cal R}$, denoted $\chi_{\cal R}$, is always larger than the entanglement entropy, and since the causal wedge is always contained within the entanglement wedge~\cite{HubRan12, Wal12, HubRanTon} (and is thus sensitive to a smaller subset of the bulk),~\cite{HubRan12} suggested that CHI is related to the von Neumann entropy by some coarse-graining procedure. Kelly and Wall proposed in~\cite{KelWal13} that CHI is dual to the ``one-point entropy'': a coarse-graining procedure of maximizing the von Neumann entropy subject to fixing the boundary one-point functions\footnote{An earlier conjecture~\cite{FreMos13} is now understood to be false~\cite{MarPC}. See~\cite{KelWal13} for details.}. Using the terminology of~\cite{Wal17}, we denote the quantity obtained by maximizing the von Neumann entropy $S$ subject to constraints $Y$ in the following way: max$(S | Y)$.


These general expectations about the dual of the causal wedge and CHI are rooted in parallels with the entanglement wedge dual. In the entanglement wedge framework, the reduced density matrix $\rho_{\cal R}=\mathrm{tr}_{{\cal R}^{c}}\rho$, where $\rho$ is the density matrix of the entire boundary, describes the dynamics of the entanglement wedge $E_{W}$~\cite{CzeKar12,Wal12, HeaHub14, JafLew15, DonHar16}, and is ignorant of the degrees of freedom outside of $E_{W}$. The von Neumann entropy of ${\cal R}$ defines the fine-grained entropy of degrees of freedom on ${\cal R}$:
\begin{equation}S_{\cal R}=-\mathrm{tr} \rho_{\cal R}\ln\rho_{\cal R},
\end{equation}
and is dual to the area of the rim of the entanglement wedge: the HRT surface~\cite{RyuTak06, HubRan07}.  From an information-theoretic perspective, the area of the HRT surface measures the uncertainty in the state $\rho_{\cal R}$ of the region $R$ given knowledge of the state of the entire boundary.


If, in accordance with with expectations (see e.g.~\cite{HubRan12, FreMos13, HubRanTon, KelWal13}), the causal wedge dual were analogous, a similar picture would hold: CHI would be the entropy of a coarse-graining that forgets everything behind the causal surface while retaining all information measurable within the causal wedge, without probing the Planck regime very close to the boundary of the causal wedge\footnote{In this hypothesis, it seems natural that the entropy associated with such a coarse-graining procedure would be proportional to the area of the causal surface, perhaps due to extensive quantum gravity degrees of freedom very near the boundary.  If these degrees of freedom were entangled with each other, their true fine-grained entropy would be smaller as measured by the boundary, consistent with the geometric fact that the HRT surface has smaller area than the causal surface.  Our results cast certain aspects of this picture into doubt, but perhaps some aspects can be salvaged}.

We find that this expectation is not borne out: if you have enough information to reconstruct the causal wedge, you already have too much information for your remaining uncertainty about ${\cal R}$ to be given by $\chi_{\cal R}$!  This demonstrates a fundamental difference between the causal and entanglement wedges, and by extension between their field theory duals. The area of the causal surface cannot be a measure of the ignorance of data inside of the causal wedge.

We present two arguments in favor of this conclusion. Our first argument relies on a rigidity property: we show in Sec.~\ref{sec:rig} that in certain situations, the causal wedge can uniquely fix the entanglement wedge, even when the former is a proper subset of the latter: $\text{Info}[C_{W}]=\text{Info}[E_{W}]$. In such cases, the data within the causal wedge fixes a unique reduced density matrix $\rho_{\cal R}$. If the causal wedge dual were parallel to the entanglement wedge dual, we would expect that whenever the entanglement wedge is rigid under a complete reconstruction of the causal wedge, $\chi_{\cal R}$ is precisely the entanglement entropy, and the two wedges coincide. However, we find that this is not generally the case. The causal wedge can fix the entanglement wedge even when it is a subset thereof, with $\chi_{\cal R}$ strictly larger than the entanglement entropy $S_{bdy}$. This is a clear contradiction to the one-point entropy conjecture, since $ {\cal S}^{(1)} = \mathrm{max}(S_{\mathrm{bdy}}|{\cal O}^{(1)}) = \mathrm{max}(S_{\mathrm{bdy}}|C_{W}) $, where ${\cal O}^{(1)}$ is the set of one-point functions, which in the classical regime is sufficient (modulo some caveats) to reconstruct the entire causal wedge, by the HKLL procedure \cite{HamKab06}. When Info$[C_{W}]=\mathrm{Info}[E_{W}]$, ${\cal S}^{(1)}=\mathrm{max}(S_{bdy}|C_{W})=\mathrm{max}(S_{bdy}|E_{W})=S_{bdy}<\chi_{\cal R}$. Thus, CHI is too coarse a quantity to be a measure of the ignorance about the general state subject to constraints sufficient to reconstruct the causal wedge, e.g. the boundary one-point functions.

Our second argument involves a quantum quench in a thermal state: by preparing a CFT state via including local sources in a Euclidean path integral, we find that even a single one-point function can constitute a constraint that is too fine to be dual to $\chi$. In particular, we find that, in our quenched system, max$(S_{bdy}|\mathcal{O})=S_{bdy}$ for $\mathcal{O}$ a single one-point function on a fixed time slice. On the bulk side, we find a contradiction: $\chi$ is strictly larger than the area of the HRT surface.

We conclude that CHI is not given by the one-point entropy.  As an added bonus, our arguments also show that CHI is not given by max$(S_{bdy}|Y)$, where $Y$ is any superset of the one-point functions; furthermore, we also rule out certain cases where $Y$ is a subset of the one-point functions.

Since the one-point functions provide an explicit reconstruction of the causal wedge, this result suggests that, if the CHI is dual to a simple boundary construction, it is not related in an obvious way to reconstruction of the causal wedge, which was part of the motivation of \cite{KelWal13}.  It might conceivably be that the causal wedge is not a distinguished surface from a boundary perspecive, but it would seem odd if holography could not naturally explain the deep connection between the area of black hole horizons and thermodynamics \cite{Haw71, Bek72, BarCar73, Haw75}.

The paper is structured as follows: in Sec.~\ref{sec:rev}, we review our definitions, the conjecture of~\cite{KelWal13} and discuss its place in a family of possible coarse graining schemes.  In Sec.~\ref{sec:overview}, we give an outline of our arguments. Sec.~\ref{sec:rig} discusses situations in which the entanglement wedge is fixed by the causal wedge and shows that the one-point entropy fails to agree with the causal holographic information, and any coarse-grained entropy obtained via constraining a superset of one-point functions will likewise fail. In Sec.~\ref{sec:therm}, we give a different type of counterexample, constructed by quenching a thermal state in the CFT; this counterexample provides an argument that fixing a subset of the one-point functions at a fixed time slice does not resolve the problem. In Sec.~\ref{sec:sum}, we summarize our arguments and conclude.

\section{Definitions of Coarse Graining} \label{sec:rev}

Let us first recall a few definitions. Consider a boundary subregion ${\cal R}$. The boundary domain of dependence $D[{\cal R}]$ of a boundary region ${\cal R}$ is the set of all boundary events that are completely determined by data on ${\cal R}$ \footnote{Formally, the domain of dependence of a region ${\cal R}$, $D[{\cal R}]$, is the union of (1) the points $p$ such that all past-directed timelike curves from $p$ (without a past endpoint) intersect ${\cal R}$ and (2) the points $q$ such that all future-directed curves from $q$ (without a future endpoint) intersect ${\cal R}$~\cite{Ger70}.}. The causal wedge of ${\cal R}$ is defined as follows (see Fig.~\ref{fig:wedges}):

\begin{equation} C_{W}[{\cal R}]\equiv I^{+}[D[{\cal R}]]\cap I^{-}[D[{\cal R}]],\end{equation}
\noindent where $I^{\pm}[D[{\cal R}]]$ are the bulk past and future of $D[{\cal R}]$.
The causal surface of ${\cal R}$, denoted $C[{\cal R}]$, is defined as the intersection of boundary of the past and future horizons of $D[{\cal R}]$:
\begin{equation}C[{\cal R}] \equiv \partial I^{+}[D[{\cal R}]]\cap \partial I^{-}[D[{\cal R}]].
\end{equation}
The causal holographic information (CHI) is defined as $\chi_{{\cal R}}\equiv \mathrm{Area}(C[{\cal R}])/4$ in Planck units.

%
The conjecture of~\cite{KelWal13} relates $\chi_{\cal R}$ to a coarse-grained entropy. By a ``coarse-grained entropy'', we mean a procedure that maximizes the von Neumann entropy subject to some constraints. A larger set of constraints corresponds to a finer coarse-grained entropy; fewer constraints result in a coarser coarse-grained entropy. In~\cite{KelWal13}, this was motivated via a connection to the Second Law. It is possible that other procedures of coarse-graining that do not involve maximizing entropy are better-suited to computing CHI; in this paper, we remain agnostic about other such procedures and treat coarse-graining only as a maximization of the von Neumann entropy.

To obtain the coarse-grained entropy proposed by~\cite{KelWal13} to be dual to CHI, we fix the one-point functions of gauge-invariant local operators supported on $D[{\cal R}]$. Denote this set, in the notation of ~\cite{KelWal13}, as $\{{\cal O}_{m}\}$. The one-point entropy ${\cal S}^{(1)}_{{\cal R}}$ associated with a boundary region ${\cal R}$ is defined as follows:
\begin{equation}
{\cal S}^{(1)}_{{\cal R}}= \mathrm{max}_{\sigma_{\cal R}}\left[ - \mathrm{Tr} \sigma_{\cal R}\ln \sigma_{\cal R}\right ] =\mathrm{max}_{\sigma_{\cal R}}S(\sigma_{\cal R})
\end{equation}
where the maximization of the von Neumann entropy $S_{\cal R}$ is over all states $\sigma_{\cal R}$ on ${\cal R}$ subject to the fixing the one-point functions of $\{{\cal O}_{m}\}$.

The conjecture proposes that under the following assumptions,
\begin{enumerate}
	\item The boundary theory is source-free (has a time-independent Hamiltonian);
	\item The bulk is a classical gravity solution of string theory (i.e. large $N$, large $\lambda$)---which presumably must obey the null energy condition;
\end{enumerate}
the one-point entropy of a boundary subregion ${\cal R}$ is dual to CHI of ${\cal R}$:
\begin{equation}
{\cal S}^{(1)}_{{\cal R}}= \chi_{\cal R}.
\end{equation}

As discussed in \cite{KelWal13}, the one-point entropy ${\cal S}^{(1)}_{{\cal R}}$ is simply one example of a broad family of coarse-grained entropies, in which one maximizes the entropy subject to the expectation values of \emph{some set} of operators.  This family of coarse grainings has a natural partial ordering with respect to ``coarseness''.  Suppose we define ${\cal S}^{(X)}$ and ${\cal S}^{(Y)}$ as the maximum entropy given the expectation values of two vector spaces $X$ and $Y$ of operators.  If $X \subset Y$, then ${\cal S}^{(X)} \ge {\cal S}^{(Y)}$ and we say that the $X$-graining is \emph{coarser} than the $Y$-graining (or equivalently, that the $Y$-graining is \emph{finer} than the $X$-graining).  This provides only a partial ordering since sometimes the sets $X$ and $Y$ might each contain observables not found in the other vector space. In this case, the two coarse-grainings are incomparable.

Since we find that CHI is not given by the one-point entropy, it is natural to wonder whether it might be dual to some other coarse-grained entropy in this class.  Our results below imply that CHI is not given by a coarser entropy than ${\cal S}^{(1)}$.  It also rules out certain classes of entropies that are finer or incomparable to ${\cal S}^{(1)}$; the details will be given in the next section.

Of course it may be that the coarse-graining procedure does not lie in the class of ``maximize S subject to some expectation values'' at all, but is actually based on totally different principles.\footnote{For example, CHI might not be a von Neumann entropy at all, but rather some combination of entropies similar to e.g. the differential entropy \cite{BalCzeCho, BalChoCze, CzeDonSul, HeaMye14, MyeRaoSug}.  Or perhaps the entropy maximization occurs over an enlarged set of possibilities which includes unphysical or inconsistent configurations.  Or maybe the coarse-graining can be identified using the information at an intermediate state in a bulk tensor network construction \cite{Swi09}.  Or it might involve restricting to some tensor factor in the Hilbert space, before calculating the entropy.}  This would be somewhat unfortunate, since using the maximization principle, \cite{KelWal13} was able to find a natural boundary explanation for the fact that the area of causal horizons is increasing in the bulk, but there might be other reasons for this area increase.


\section{Overview of Arguments}\label{sec:overview}
Here we provide a sketch of our arguments. The reader who is not concerned with the details may safely skip Sections~\ref{sec:rig} and~\ref{sec:therm} after reading this section.

In the subsequent section, we will argue that  any set of constraints sufficient to fully reconstruct the causal wedge has insufficient uncertainty to yield an entropy that can be dual to CHI. One consequence is that the one-point entropy conjecture above is not sufficiently coarse-grained to be dual to CHI. The argument consists of a logical sequence of three steps:
\begin{enumerate}
	\item Consider a pure bulk geometry for which the one-point boundary data for ${\cal R}=\partial M$ is sufficient to reconstruct the entire entanglement wedge (we will construct an explicit example of such a spacetime in Sec.~\ref{sec:rig}): Info$[C_{W}]=\mathrm{Info}[E_{W}]$.
	\item It immediately follows from the previous point that $\sigma$ is pure, and thus the one-point entropy vanishes: ${\cal S}^{(1)}=0$.
	\item The causal wedge is (in the generic states that we construct) a proper subset of the entanglement wedge, so that CHI is equal to a strictly positive area: $\chi>0$.
\end{enumerate}
These may be summarized by an equation, where we use the notation $\stackrel{\mathrm{eg}}{=}$ ($\stackrel{\mathrm{eg}}{\neq}$) to denote an equality (inequality) that holds in our specific examples:

\begin{equation}\mathrm{max}(S_{\mathrm{bdy}}|C_{W}) \stackrel{\mathrm{eg}}{=} \mathrm{max}(S_{\mathrm{bdy}}|E_{W}) = \mathrm{Area}[X]\stackrel{\mathrm{eg}}{\neq} \mathrm{Area}[C],
\end{equation}
where $X$ and $C$ are HRT and causal surfaces, respectively.

We immediately arrive at a contradiction. This shows that not only ${\cal S}^{(1)}$, but also any finer proposal (maximizing $S$ subject to a superset of the one-point constraints) fails to be dual to CHI.  Not only that, but any set of constraints sufficient to reconstruct the causal wedge, is already too fine to be correct. As discussed in the Introduction, this is very surprising as it is not parallel to the case of entanglement wedge reconstruction: $\chi_{\cal R}$ cannot be an entropy associated to the degrees of freedom in the causal wedge, as $S_{\cal R}$ is for the entanglement wedge.

In Sec.~\ref{sec:therm}, we argue that a set of constraints \textit{coarser} than the one-point functions of the boundary at one moment in time also fails to yield a coarse-grained entropy dual to CHI. To do this, we consider a state that maximizes the von Neumann entropy subject to fixing some particular one-point operator $H + \cal{O}$ at $t = 0$.  These states may be constructed at $t = 0$ by means of a Euclidean path integral on a thermal imaginary time circle with a static source added to the normal Hamiltonian time evolution.  However, in order to satisfy the ``no sources'' criterion of \cite{KelWal13}, we turn this source off when considering the Lorentzian time evolution of the state to the future or past of $t = 0$.  This results in a ``quenched'' state \cite{CalCar06}.  This procedure does not always give rise to a well-defined state~\cite{VanU}, but when it does, in such states generically ${\cal S}^{(1)} < \chi$.

In addition to forbidding coarse-graining procedures that are finer than the one-point entropy, this counterexample also rules out certain coarser proposals, in which only a subset of the one-point functions are constrained---so long as constraining the $t = 0$ subset leads to a well defined state.  It further forbids any proposal coarser than these.  Of course, if a coarse-grained state is \emph{not} well defined, then its entropy also cannot be dual to CHI.

This leaves only the following possibilities for the correct boundary dual to $\chi$:
\begin{enumerate}
\item Coarse grain by a set of constraints that is ``incomparable'' to the one-point entropy (neither a subset, nor a superset);
\item Coarse grain by a subset of one-point functions that lead to a well-defined state $\sigma$ when you constrain them on the entire wedge $D[R]$, but not when you only constrain them at $t = 0$;
\item Some notion of ``coarse graining'' that does not require maximizing the von Neumann entropy subject to a set of constraints. This includes the possibility of a complicated quantity with no clear information-theoretic interpretation.
\end{enumerate}
Except in the last case, the constraints must also be insufficient to reconstruct the causal wedge, as we have said.

\section[Enough Constraints are Too Many: A Rigid Interior]{Enough Constraints are Too Many: A Rigid{\linebreak} Interior}\label{sec:rig}

In this section, we give an example in which a complete description of the causal wedge of $\partial M$ is sufficient to uniquely fix the entanglement wedge $E_{W}$ of $\partial M$. Using this construction, we argue as outlined above that CHI is not dual to any coarse-graining that is constrained by data sufficient to fix the entanglement wedge. This also implies that in such cases, CHI generically does not coincide with the one-point entropy ${\cal S}^{(1)}$.

We now fill in the details of the argument outlined in the previous section. If $C_{W}[\partial M]$ uniquely fixes $E_{W}[\partial M]$, then there is a unique boundary density matrix $\rho$ corresponding to $C_{W}[\partial M]$. This last point follows from the proof of~\cite{DonHar16} that the reduced density matrix $\rho_{\cal R}$ of a boundary subregion ${\cal R}$ is dual to entanglement wedge of ${\cal R}$, $E_{W}[{\cal R}]$. Thus if $C_{W}[\partial M]$ fixes $E_{W}[\partial M]$, it immediately follows that the causal wedge also fixes the density matrix $\rho$. Any set of constraints that specifies the causal wedge geometry thus also specify the full density matrix, and coarse-graining the von Neumann entropy subject to fixing the density matrix trivially yields the von Neumann entropy of the same density matrix. So we find that ${\cal S}^{(1)}=S=0$. Generically, however, $\chi>0$.

In the particular case of the one-point entropy, the set of constraints is the set of one-point functions. Following the procedure of HKLL~\cite{HamKab06}, $C_{W}[\partial M]$ can be reconstructed for any classical geometry from the one-point functions (and the field equations): thus in this example there is a unique state $\rho$  with the specified set of one-point functions. This immediately implies that the one-point entropy vanishes; however, when $C_{W}[\partial M]$ is a proper subset of $E_{W}[\partial M]$, as is the case whenever null geodesics encounter any curvature, the area of $C[\partial M]$ is nonzero, and thus cannot be equal to the one-point entropy.

Let us fortify this argument with an explicit example, which also serves as justification of the assumptions above for the reader who might otherwise be skeptical that the causal wedge can uniquely fix the entanglement wedge when the former is a proper subset of the latter.

\begin{figure}[t]
\centering
\includegraphics[width=8cm]{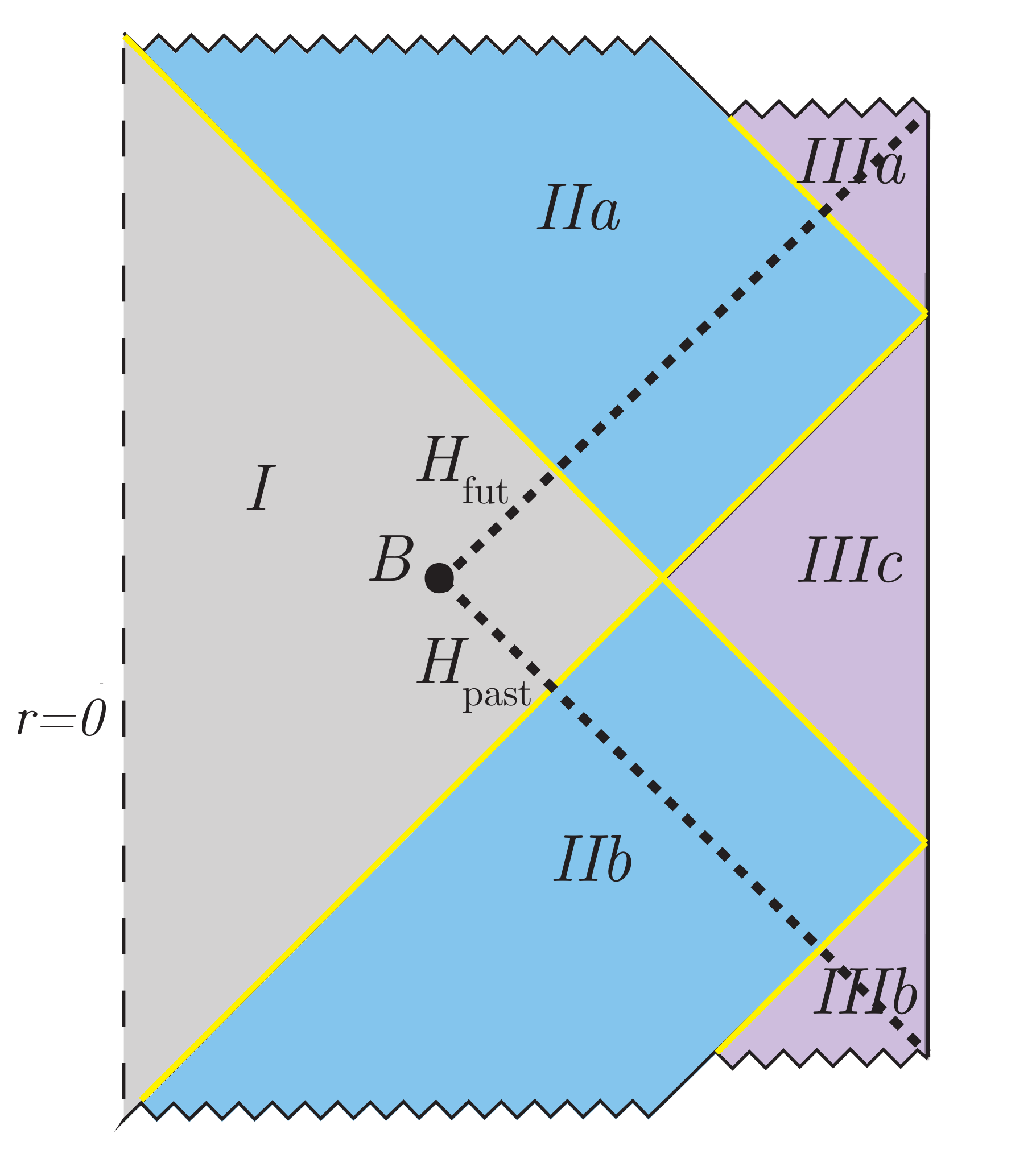}
\caption{The cut-and-paste geometry is constructed from six patches of different mass AdS-Schwarzschild. These regions are patched together via AdS-Vaidya in the thin-shell limit. The bifurcation surface $B$ is the causal surface of the entire boundary. It lies well within the vacuum AdS ($M=0$) region of the spacetime.}
\label{fig:cutandpaste}
\end{figure}

We consider firing two identical time-symmetric null shells into vacuum AdS. Since the shells bounce off of the boundary, the boundary theory is source-free. The effect of one shell creates black and white holes of mass $M$, with future and past event horizons. See Fig.~\ref{fig:cutandpaste}. A second null shell brings the white and black hole singularities closer together, shifting the event horizons to intersect on a nontrivial bifurcation surface $B$.\footnote{One can also obtain a nontrivial bifurcation surface by firing a single shell with sufficient mass to ensure that it collapses into its Schwarzschild radius sufficiently quickly.}.

\begin{figure}[t]
\centering
\subfigure[ ]{
\centering
\includegraphics[width=0.26 \textwidth]{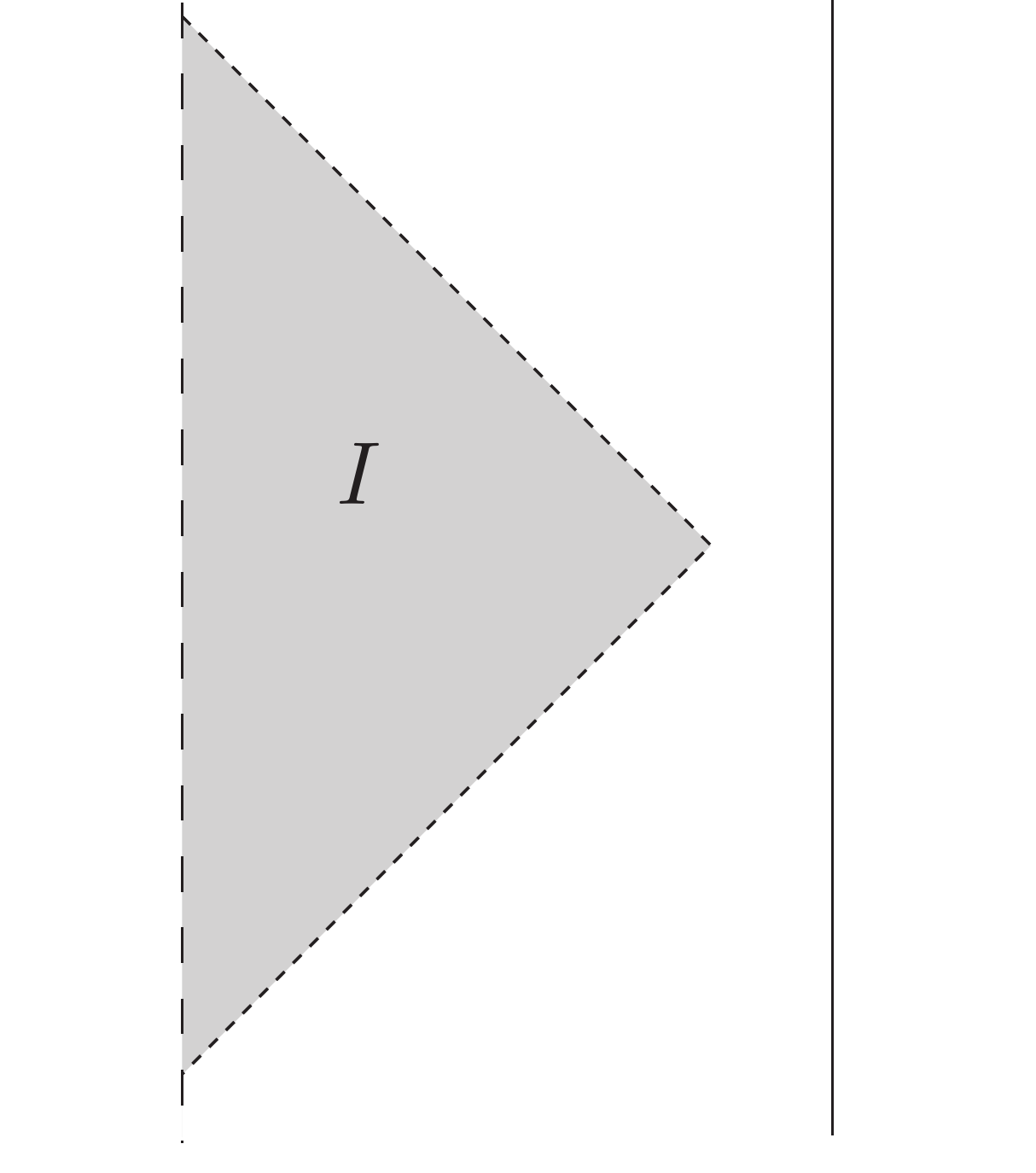}
\label{fig:RegionI}}
\qquad
\subfigure[ ]{
\centering
\includegraphics[width=0.26 \textwidth]{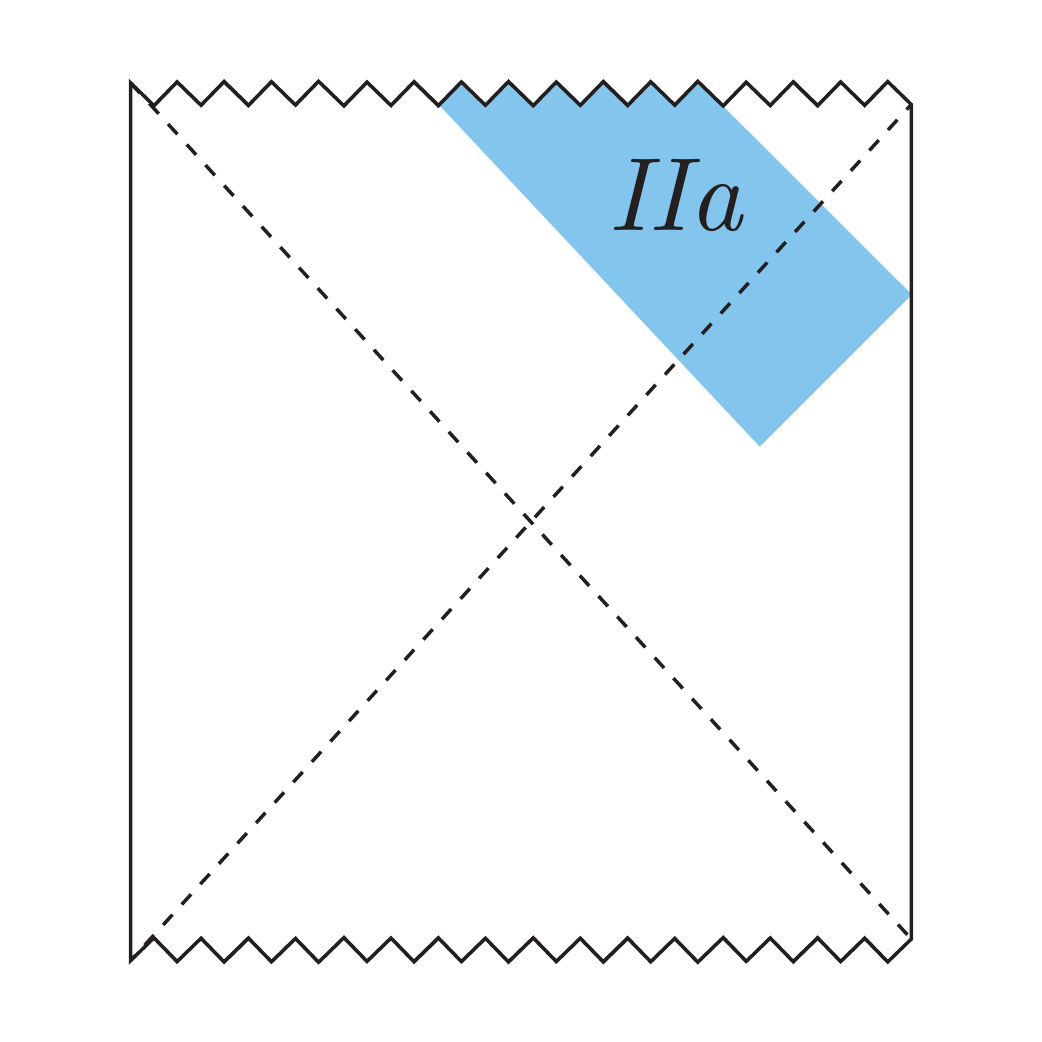}
\label{fig:RegionIIa}}
\qquad
\subfigure[]{
\centering \includegraphics[width=0.26\textwidth]{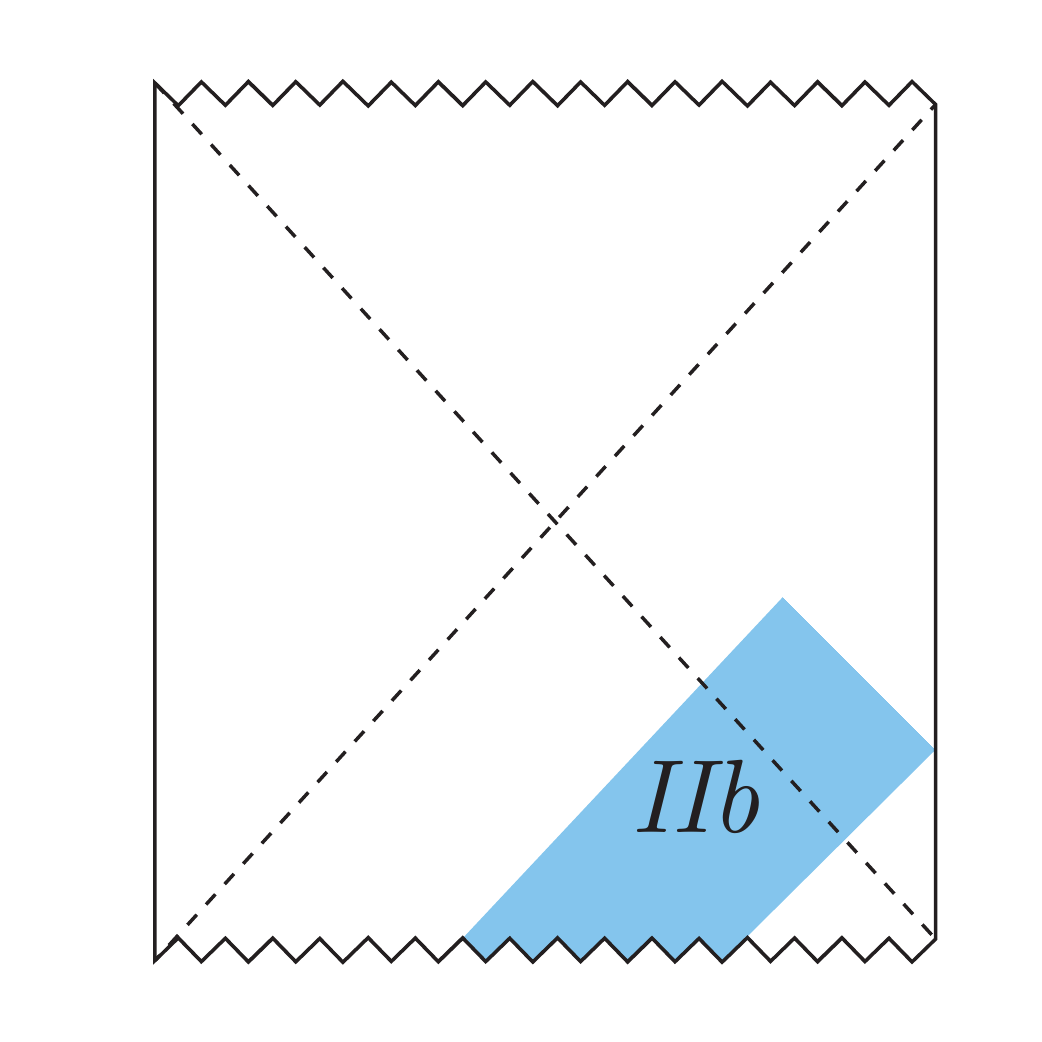}
\label{fig:RegionIIb}}
\qquad
\subfigure[]{
\centering \includegraphics[width=0.26\textwidth]{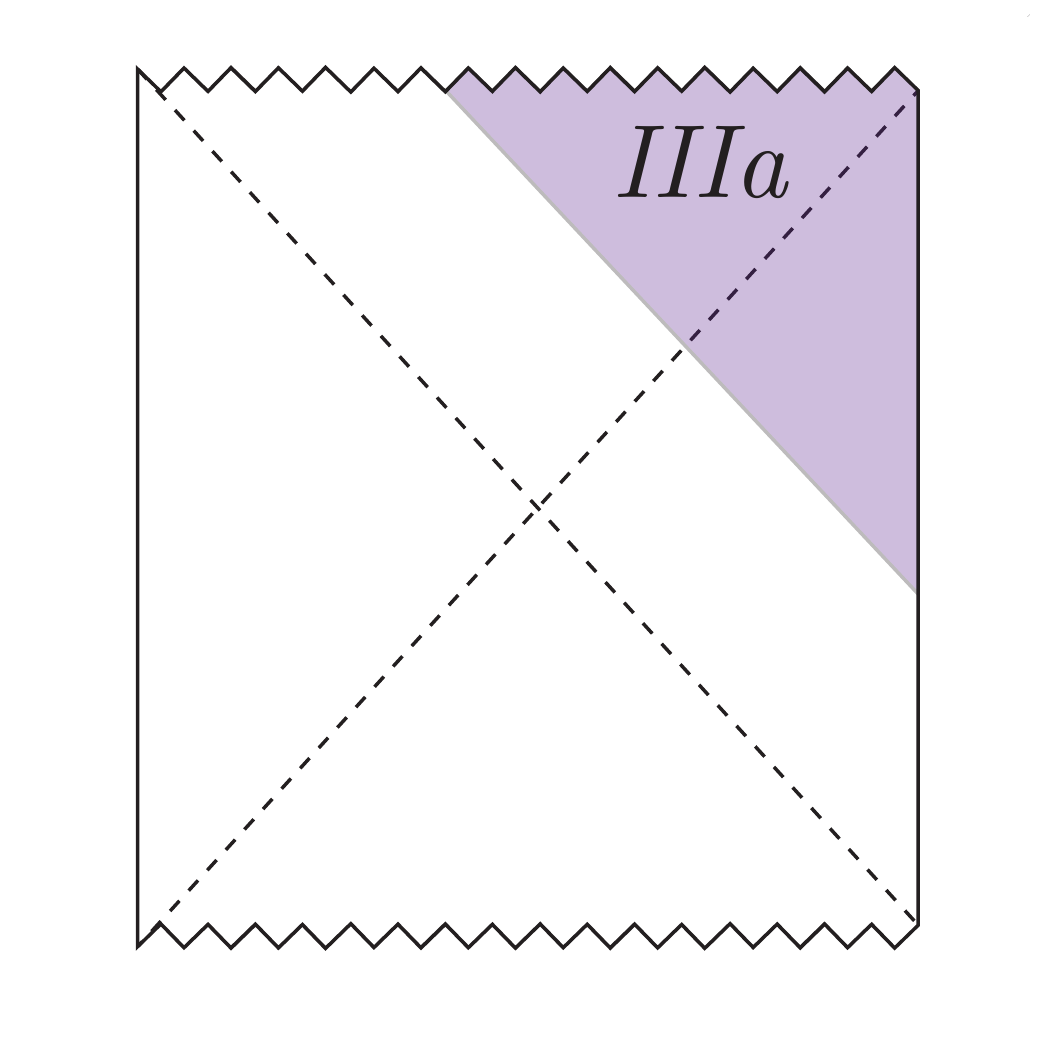}
\label{fig:RegionIIIa}}
\qquad
\subfigure[]{
\centering \includegraphics[width=0.26\textwidth]{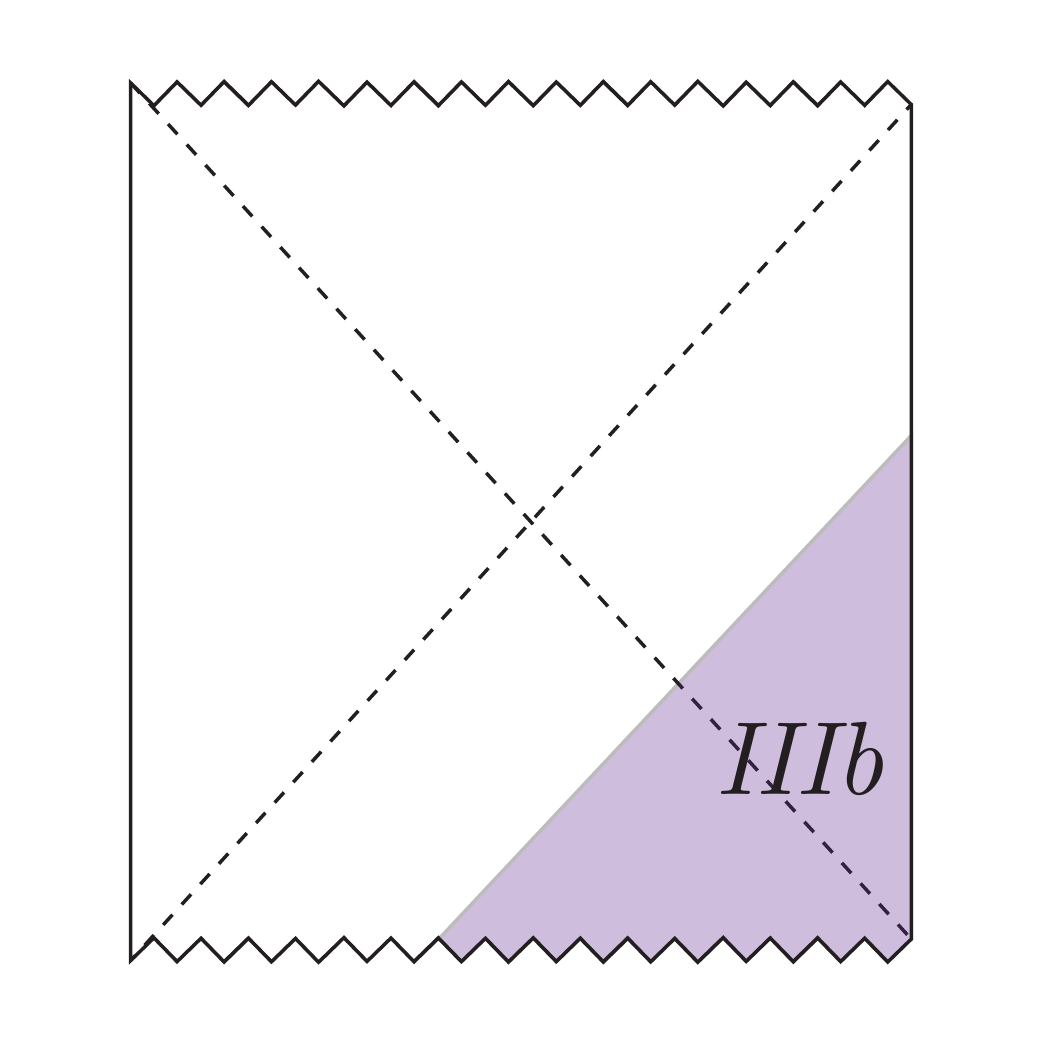}
\label{fig:RegionIIIb}}
\qquad
\subfigure[]{
\centering \includegraphics[width=0.26\textwidth]{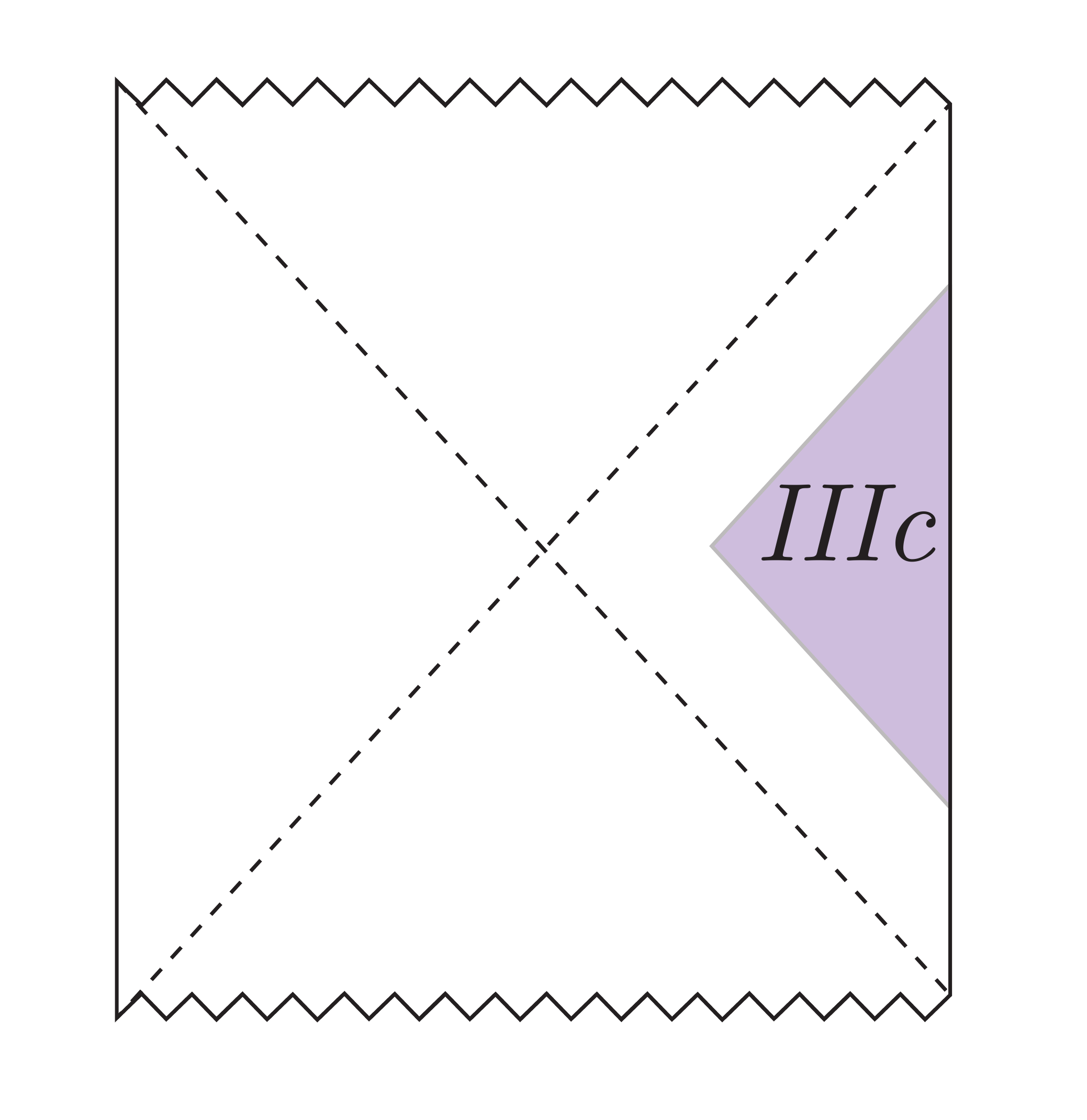}
\label{fig:RegionIIIc}}
\caption{(a) A patch of AdS (zero mass Schwarzschild, in more than 3 bulk dimensions. (b)(c) Patches of mass $M$ AdS-Schwarzschild. (d)(e)(f) Patches of mass $2M$ AdS-Schwarzschild.}
\label{fig-patches}
\end{figure}

More precisely, the spacetime in question, illustrated in Fig.~\ref{fig:cutandpaste}, is a simple cut-and-paste construction similar to that of~\cite{FisMar14}. Region I is the patch of pure AdS illustrated in Fig.~\ref{fig:RegionI}, Regions IIa and IIb are patches of a Schwarzschild-AdS black hole of mass $M$, see Figs.~\ref{fig:RegionIIa},~\ref{fig:RegionIIb}. Regions IIIa-c correspond to patches of a Schwarzschild-AdS black hole of mass $2M$, Figs.~\ref{fig:RegionIIIa},~\ref{fig:RegionIIIb},~\ref{fig:RegionIIIc}.

This geometry may be modeled via six patches of AdS-Schwarzschild with different masses, patched together via ingoing and outgoing AdS-Vaidya shells in the thin-shell limit. The ingoing and outgoing metrics may be written in coordinate form, where $v$ and $u$ are ingoing and outgoing null coordinates:
\begin{align}& ds^{2}_{in} = -f(r,v)dv^{2} +2 dv dr +r^{2} d\Omega_{d-1}^{2}\\
& ds^{2}_{out} = -f(r,u) du^{2} -2dudr +r^{2} d\Omega_{d-1}^{2},
\end{align}
where $f(r,w) = 1- M(w)r^{2-d}+r^{2}$, for $w=u$ or $v$, $d\Omega_{d-1}$ is the metric of a ($d-1$)-sphere ($d$ being the boundary spacetime dimension) and we have fixed the AdS length scale to 1. These are asymptotically AdS solutions to the Einstein Field Equation with stress tensors:
\begin{align}
&T_{ab}^{(in)} \propto \frac{dM(v)}{dv} (\partial_{v})_{a} (\partial_{v})_{b}\\
&T_{ab}^{(out)} \propto- \frac{dM(u)}{du} (\partial_{u})_{a} (\partial_{u})_{b}.
\end{align}
Picking $M'(v)\geq 0$ and $M'(u)\leq 0$ yields a geometry that obeys all standard energy conditions. Since the shells are identical, the resulting spacetime contains patches of mass $0, M$, and $2M$. By taking the time difference at which the shells are fired to be sufficiently large, we are guaranteed that the past and future horizons intersect at a nontrivial bifurcation surface $B=C[\partial M]$ of nonzero area.

Note now that the bifurcation surface lives well inside the vacuum AdS region of the geometry, and that the interior of $B$ on any AdS-Cauchy slice (see~\cite{EngWal14} for a definition of AdS-Cauchy) also lives in vacuum AdS. Positivity of energy in the boundary field theory implies positivity of energy in the bulk; therefore one expects that AdS is the ground state of the bulk system.  Assuming the boundary CFT has a unique ground state, the bulk vacuum is likewise rigid:\footnote{One can also prove bulk rigidity via the saturation condition of the Positive Energy theorem~\cite{Wit81}, but only for theories with matter that obeys the Dominant Energy Condition, which is known to be violated in AdS by tachyonic scalars even if they obey the Breitenlohner-Freedman bound~\cite{BreFre82} and the Null Energy Condition.  Presumably an extension of the Positive Energy Theorem covers this situation, but to the best of our knowledge nobody has yet shown this.} the interior of $B$ cannot be replaced by a different connected\footnote{The restriction to connected geometries is to disallow tensoring AdS with a closed cosmology.  It is unclear how the holographic paradigm should be extended to closed universes, but presumably any information in such a universe cannot be reconstructed from the boundary CFT \cite{MarWal12}, and is therefore not relevant for our purposes.} spacetime region.

Since $B$ is the boundary of the causal wedge $C_{W}[\partial M]$, there is a unique bulk geometry corresponding to $C_{W}[\partial M]$: $C_{W}[\partial M]$ fixes the data on a complete AdS-Cauchy slice. There is then a unique boundary state $\rho$ corresponding to the \textit{causal} wedge $C_{W}[\partial M]$, which can be reconstructed from one-point functions. The one-point entropy ${\cal S}^{(1)}$  must therefore vanish by the purity of $\rho$, in contradiction with the conjecture that ${\cal S}^{(1)}= \chi=\mathrm{Area}(C[\partial M])/4\neq 0$.

In this example, the set of one-point functions is not sufficiently coarse to be used in constructing a dual to CHI. Any superset of the boundary one-point functions is even finer, and thus also cannot be dual to CHI. Similarly, a different set of constraints sufficient to fix the causal wedge is also wrong.

Note that our argument can be applied to the regime of free semiclassical fields if one constrains the pulses to be a bulk coherent state.  Such states minimize the energy subject to the the expectation value of the bulk field, and are therefore still uniquely specified by the one-point data on the boundary.

\section{Fewer Constraints are Still Too Many: Thermal Quenches} \label{sec:therm}

In light of the results above, a natural attempt to correct the one-point entropy conjecture is then to use as a constraint a subset of the one-point functions rather than the full set. In this section, we argue that fixing a smaller subset of one-point functions -- or even a single one -- on a constant time slice still does not yield the causal holographic information.

Recall the standard construction of the thermofield double dual to the eternal Schwarzschild-AdS black hole. The CFT state is prepared via a Euclidean path integral on an imaginary thermal circle. The CFT is on $S^{1}\times S^{d-1}$ with Hamiltonian $H$ generating evolution in Euclidean time $\tau$; the interior is a smooth geometry that analytically continues to Schwarzschild-AdS (since the geometry is classical, we are above the Hawking-Page phase transition~\cite{HawPag83}) when $\tau$ is continued to Lorentzian time $t=i\tau$ from $\tau=0$.

\begin{figure}[t]
\centering
\includegraphics[width=8cm]{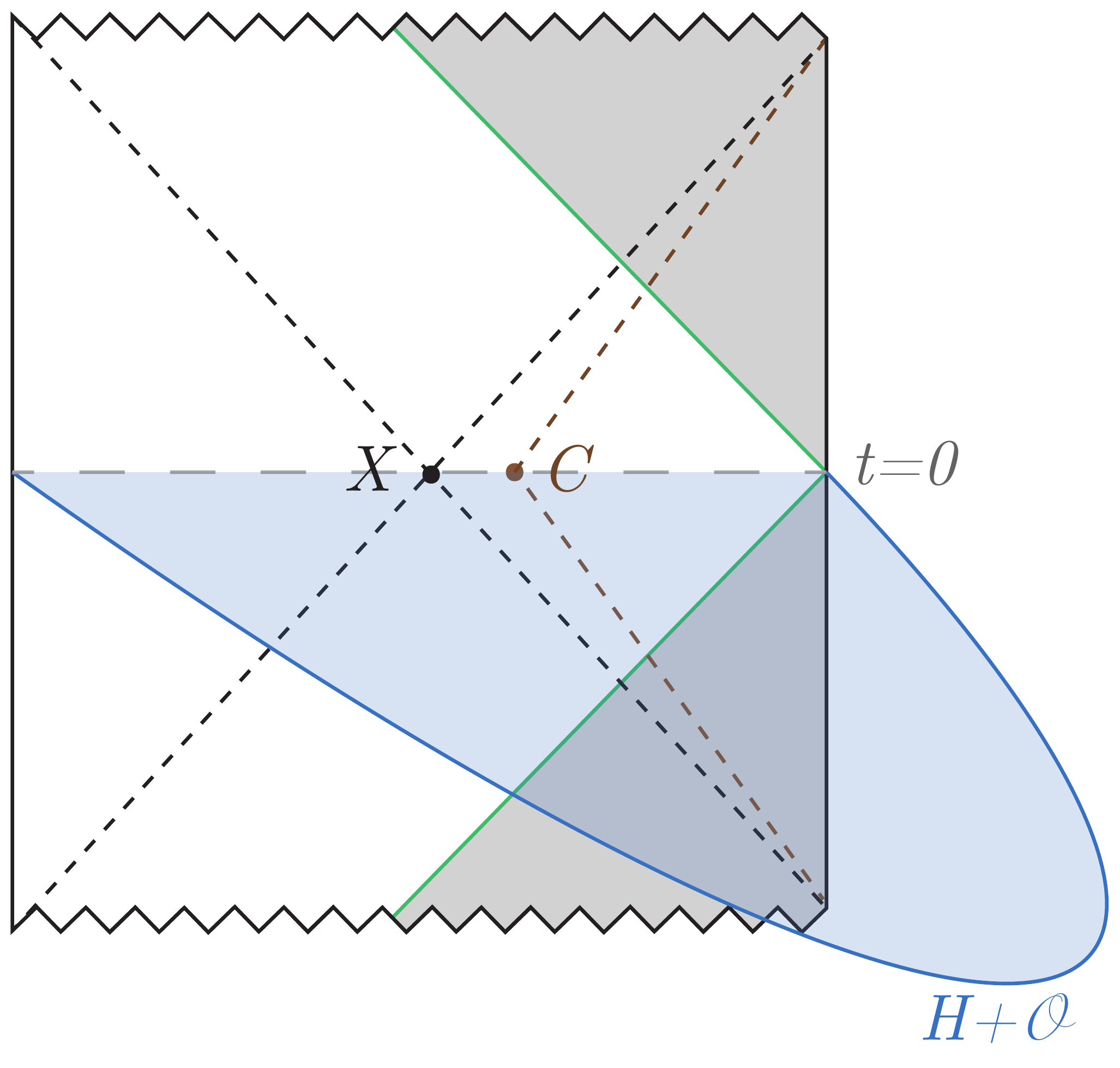}
\caption{The two geometries, corresponding to an unquenched (AdS-Schwarzschild) geometry, and the quenched geometries, are superposed. The extremal surface $X$ is unaltered by the matter, while the causal surface $C$ is displaced.  The areas in gray represent the region that is perturbed by matter, and the green lines represent its wavefront (whose maximum velocity is the speed of light). The Euclidean geometry is drawn in blue as an additional branch, coinciding with the Lorentzian geometry at $t=0$. (Because
of the possibility of caustics, the null surfaces coming from $X$ and $C$ are drawn with different angles, even though both surfaces are generated by null geodesics.}
\label{fig:quench}
\end{figure}

In a departure from the usual construction, we will add one or more static, locally-sourced operators $\mathcal{O}$ to the Euclidean path integral on the thermal circle: $H\rightarrow H+ \mathcal{O}$.  This path integral produces a grand canonical ensemble where the source is the chemical potential for $\mathcal{O}$.  In order to accommodate the requirement of~\cite{KelWal13} that the Lorentzian theory be source-free, we turn the sources $\mathcal{O}$ off at $\tau=0$, resulting in a quenched state~\cite{CalCar06} (Fig.~\ref{fig:quench}).  If this construction succeeds, we now have a state $\rho(t = 0) = \exp[\beta(H + \mathcal{O})]$ which maximizes the von Neumann entropy at $t = 0$, while fixing the particular one-point function $\langle H + \mathcal{O}\rangle$ at time $t = 0$.

For some choices of $\mathcal{O}$, this construction might not give a well-defined state~\cite{VanU}.  This can happen in a couple of different ways.  First, if $\mathcal{O}$ has dimension $\Delta > d$ exceeding the dimension of the boundary spacetime, its source corresponds to an irrelevant coupling and the boundary theory might not be defined in the UV\footnote{When $\Delta = d$, a nonlinear analysis of the RG flow is needed to determine if the coupling is marginally relevant or irrelevant.}.  Secondly, if the spectrum of $H + \mathcal{O}$ is not bounded below, or if it is Hagedorn for the choice of $\beta$, then the thermal state will not be well-defined.  Third, it may be that the state is well-defined when the source for $\cal{O}$ is turned on, but that the instantaneous quench is too abrupt and one ends up in a state with infinite energy.  In any of these situations, we cannot expect that $\rho$ will have a well-defined bulk dual\footnote{For higher dimensional operators, it might be possible to regulate the instantaneous quench in the UV by smoothing it out over some small but finite time. Since we need the boundary theory to be source-free, the regulation would need to happen in the Euclidean time, while still allowing us to fix the one-point functions.  This would complicate our argument below, but we expect that the end result would still be the same.}.

However, there are known examples of $\cal{O}$'s that give rise to well-defined instantaneous quench states (e.g. \cite{CalCar06, CalCar07, SotCar10}, and in the holographic context see e.g. \cite{Das11,LiuSuh13a, LiuSuh13b} to name but a few), and we expect that the quench is well defined whenever $\cal{O}$ is a sufficiently low dimension operator.  For scalar sources, it has been argued (for holographic \cite{BucMye13} and general \cite{DasGal14,Das:2014jna,Das:2015jka} quenches) that the expected final energy $\langle H \rangle$ converges iff
\be\label{bound}
\Delta < d/2,
\ee
corresponding to an alternately quantized scalar field in AdS/CF.\footnote{Even when $\langle H \rangle$ diverges, the state should still have support on finite $H$ when $\Delta < (d+2)/2$.  This bound can be derived by requiring all divergences in the effective action $-\ln Z$ to be integrable across the quench.}.  To see that $\langle H \rangle = \text{finite}$ is the correct criterion for our argument, note that in what follows we will be interested in the areas of certain bulk surfaces.  The area-entropy of a thermal black hole is related to the CFT energy by the standard blackbody formula $A \sim H^{\frac{d-1}{d}}$, while perturbations to the horizon due to a bulk stress-tensor $T_{ab}$ which we expect to be of the same order as $H$.  Hence, finiteness of $H$ should guarantee that our geometrical constructions involve convergent quantities\footnote{In fact, if we only demand finiteness of $H^{\frac{d-1}{d}}$ (which bounds above the area of the final black hole and hence the areas of $X$ and $C$), one might be able to get away with slightly higher values of $\Delta$.}.

The existence of even one field satisfying \eqref{bound} is sufficient to produce a counterexample to the one-point entropy conjecture, as well as to any conjecture in which one maximizes subject specifically to the operator $H + \mathcal{O}$ at $t=0$ (irrespective of whether there are any additional constraints).

We now consider the case in which the quenched state is well-defined, and consider its bulk dual.  Had we not performed the quench, our state would simply be a stationary thermal state, and will therefore be dual to a smooth stationary black hole, i.e. a bifurcate Killing manifold. In this geometry, both the causal surface $C$ and the HRT surface $X$ are simply the Killing bifurcation surface, and thus $\chi=S$.

The effect of the local source $\mathcal{O}$, which is turned off at $t=0$, is to create some additional positive-energy matter propagating into the solution in both time directions, with retarded modes coming from $t > 0$ and advanced modes coming from $t < 0$.  The wavefront of the perturbation propagates in at the speed of light, and bouncing off of $t = 0$ as shown in Fig.~\ref{fig:quench}.  The geometry of the region inside the wavefront is unaffected by the quench, so $X$ and therefore $S$ remain unaltered.  However, the causal surface $\chi$ is defined teleologically, and in the classical limit it generically increases in area as a result of matter falling across the horizons \cite{Wal12}.    Hence $\chi>S$.

Recall now that the state $\rho$ was constructed via a Euclidean path integral to maximize $S$ given its value of $\left \langle H+{\cal O}\right\rangle$, where ${\cal O}$ is an integral of some one-point data.  Since fixing a subset of the one-point functions at $t=0$ uniquely specifies the state $\rho$, we find $S=S^{(1, t=0)}$, the entropy coarse-grained entropy subject to fixing one-point functions at $t=0$. Since including one-point functions at later or earlier times can only result in a finer quantity, ${\cal S}^{(1,t=0)}\geq {\cal S}^{(1)}$. Altogether:
\begin{align}S=&{\cal S}^{(1,t=0)} \geq {\cal S}^{(1)}\geq S\\
& \therefore {\cal S}^{(1)}=S.
\end{align}
However, $\chi>S$, so we derive a contradiction with $\chi={\cal S}^{(1)}$. As this contradiction is obtained even if we constrain just a single one-point function, it is clear that not only does the one-point entropy conjecture fail, but any coarser variant involving only a subset of the one-point functions must fail as well. We have thus found that, paradoxically, the one-point entropy conjecture is simultaneously too fine and too coarse! But recall that the space of coarse-grainings is partially ordered with respect to coarseness, not totally ordered. The set of data sufficient for reconstruction of the causal wedge is \textit{incomparable} with the constraints over which the dual to $\chi$ is coarse-grained, if $\chi$ is indeed obtained by a coarse-graining procedure involving a maximization subject to constraints. As noted in Sec.~\ref{sec:rev}, it is possible an altogether different type of coarse-graining yields the correct dual to the causal wedge.

\paragraph{Quantum Corrections.} A similar argument can be made when the bulk receives perturbative quantum corrections. By perturbative quantum corrections, we mean that the bulk geometry admits a perturbative expansion in $G\hbar$ around a classical background. To first order in $G\hbar$, the entanglement entropy of a boundary subregion ${\cal R}$ is given by~\cite{FauLew13}
\begin{equation}
S_{\cal R} = \frac{\mathrm{Area}(X_{\cal R})}{4} +S_{\mathrm{ent}}(X_{\cal R})+\mathrm{counterterms}\equiv S_{\mathrm{gen}}(X_{\cal R}),
\end{equation}
where $X_{\cal R}$ is the HRT surface, $S_{\mathrm{ent}}(X_{\cal R})$ is the bulk entanglement entropy across $X_{\cal R}$ due to bulk quantum fields. At higher orders in perturbation theory, we proposed in~\cite{EngWal14} that $S_{\cal R}$ is given by:
\begin{equation}
S_{\cal R}=S_{\mathrm{gen}}(\varkappa_{\cal R}),
\end{equation}
where $\varkappa_{\cal R}$, called the quantum extremal surface, extremizes $S_{\mathrm{gen}}$ rather than just the area. We showed in~\cite{EngWal14} that, assuming that the bulk obeys the Generalized Second Law~\cite{Bek73, Haw76}, $\varkappa_{\cal R}$ lies deeper than (and in a spacelike direction to) the causal surface $C[{\cal R}]$. Assuming the Generalized Second Law, the arguments above carry over to the perturbatively quantum bulk regime, under the substitution of the causal holographic information with a ``quantum'' causal holographic information: $S_{\mathrm{gen}}(C[{\cal R}])$, rather than the area of $C[{\cal R}]$.

\paragraph{Nonlocal Constraints.} On a somewhat more aggressive note, we can also consider the effects of maximizing $S$ subject to some set of operators $\cal{O}$ which are not fully local in space.  If the Hamiltonian $H + \cal{O}$ is completely nonlocal, then there is no good reason to expect the corresponding bulk field theory to be local either.  This invalidates our argument that the geometry remains a smooth black hole after the quench.  However, if $\cal{O}$ is only mildly nonlocal, for example, if it is linear in data that is localized to small intervals, then one might expect the nonlocal aspects of the quench to perturb the state only close to the conformal boundary.  Assuming that $H + \cal{O}$ flows under renormalization to a local Hamiltonian in the IR, and that this local Hamiltonian is still holographically dual to the same bulk gravity model,\footnote{This is guaranteed if $\cal{O}$ is only a small perturbation to the holographic Hamiltonian $H$, or failing that if it only contributes to irrelevant terms so that it is small in the IR.} all of the same arguments can still be made, in particular that $\chi > S$ so that the coarse-graining is not CHI.

This allows one to extend the argument to e.g. one-point data in a small but finite duration time strip $V$ near $t = 0$.  One simply uses boundary time evolution to view local operators in $V$ as slightly nonlocal operators at $t = 0$.  This helps to alleviate worries about the need to smear certain operators in time, and extends our counterexample to even more kinds of coarse-graining.

\section{Conclusions}\label{sec:sum}

We have argued that (breaking with intuitions guided by the behavior of entanglement-based bulk reconstruction) the area of the causal surface cannot be obtained via a coarse-graining procedure that maximizes the von Neumann entropy subject to any set of constraints that fixes the causal wedge.

We described two counterexamples in Sections \ref{sec:rig} and \ref{sec:therm} that rule out the one-point entropy and various related coarse-graining schemes.  The counterexamples can be constructed in the regime where the bulk is classical, but they can also be extended to geometries with perturbative quantum corrections, as discussed towards the end of these sections.

\begin{figure}[t]
\centering
\includegraphics[width=12cm]{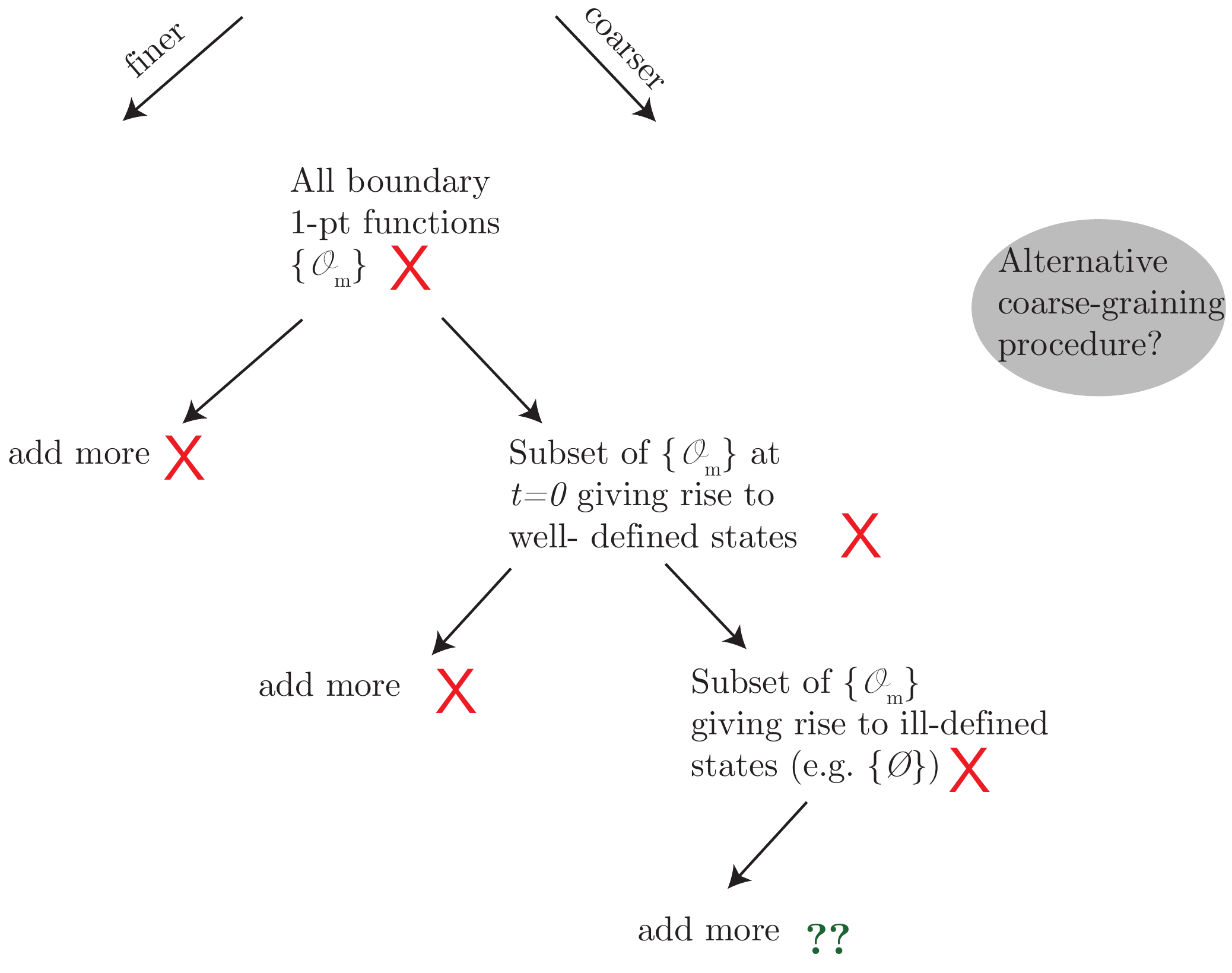}
\caption{An illustration of all possible constraints for a coarse-grained entropy dual to CHI. We have ruled out a maximization coarse-graining procedure constrained by (1) the set of all boundary one-point functions, (2) any superset of the boundary one-point functions, (3) any subset of the one-point functions giving rise to well-defined states, (4) any superset of (3), and (5) any subset of the one-point functions giving rise to an ill-defined state. The remaining options are supersets of (5), or alternative coarse-graining procedures not based on entropy maximization.}
\label{fig:constraints}
\end{figure}

The coarse-graining proposals we have ruled out are maximizing the entropy subject to any of the following:
\begin{enumerate}
\item any set of constraints sufficient to reconstruct the causal wedge,
\item the one-point functions (which are sufficient to reconstruct classically),
\item any superset of the one-point functions,
\item any subset of the one-point functions at $t = 0$ whose grand canonical ensemble is a well-defined thermal state,
\item any superset of a subset in the previous category,
\item and of course any set whose whose grand canonical ensemble is ill-defined.
\end{enumerate}
Note that our first counterexample in \ref{sec:rig} addresses 1-3, and the second counterexample in \ref{sec:therm} addresses 2-6.  We have not ruled out supersets of those subsets giving rise to ill-defined states (doing so would be difficult, as the empty set is one such subset, and is a subset of all sets!).  Nor have we ruled out the possibility that a subset of the one-point functions not restricted to $t = 0$ might be the correct choice, but this is possible only if the $t = 0$ part of the constraint does not give rise to a well-defined state.  Finally, we have not proven that the ``maximize entropy subject to constraints'' paradigm is the correct way to think about the CHI at all---indeed our results place significant pressure on this picture!  The situation is summarized in Fig.~\ref{fig:constraints}, which shows the relationship of these coarse-grainings according to the partial ordering defined in \ref{sec:rev}.

As discussed in Sec.~\ref{sec:rev}, these results demonstrate a basic difference between the causal and entanglement wedges, and presumably by extension, between the causal and entanglement reconstruction schemes.  In light of recent progress on both fronts, it is intriguing and likely instructive that the two approaches are divergent.

This observation also raises the possibility that, unlike the area of the HRT surface, CHI simply has no simple information-theoretic dual.
\\

\end{spacing}

\noindent \textbf{Acknowlegements:} It is a pleasure to thank Sebastian Fischetti, Will Kelly, Don Marolf, Mark Mezei, Douglas Stanford, and Mark Van Raamsdonk for helpful discussions. The work of NE was supported in part by NSF grant PHY-1620059. The work of AW was supported by the Institute for Advanced Study, the Martin A. and Helen Chooljian Membership Fund, the Raymond and Beverly Sackler Foundation Fund, and NSF grant PHY-1314311.

\bibliographystyle{JHEP}

\bibliography{all}

\end{document}